\documentclass[journal]{IEEEtran}

\usepackage{cite}

\usepackage[pdftex]{graphicx}
\usepackage{rotating}
\usepackage{amsmath}

\usepackage{algorithm}
\usepackage{algorithmic}

\usepackage{multirow}

\usepackage{cite}
\ifCLASSINFOpdf

\else

\fi

\usepackage{fixltx2e}

\begin{document}
	
	\title{Deep Learning Head Model for Real-time Estimation of Entire Brain Deformation in Concussion}

	\author{Xianghao~Zhan,
		Yuzhe~Liu,
		Samuel~J.~Raymond,~\IEEEmembership{Member,~IEEE,}
		Hossein~Vahid~Alizadeh,~\IEEEmembership{Member,~IEEE,}
		August~G.~Domel,
		Olivier~Gevaert,
		Michael~Zeineh,
		Gerald~Grant,
		and~David~B.~Camarillo,~\IEEEmembership{Member,~IEEE}
		\thanks{X. Zhan, Y. Liu, S. Raymond, H. Vahid Alizadeh, A. Domel, and D. Camarillo are with the Department of Bioengineering, Stanford University, Stanford, 94305, USA. X. Zhan and Y. Liu contributed equally to this work. (Corresponding Author: D. Camarillo e-mail: dcamarillo@stanford.edu)}
		\thanks{O. Gevaert is with the Department of Biomedical Informatics, Stanford University, Stanford, 94305, USA}
		\thanks{M. Zeineh is with the Department of Radiology, Stanford University, Stanford, 94305, USA}
		\thanks{G. Grant is with the Department of Neurosurgery, Stanford University, Stanford, 94305, USA}
		\thanks{This work has been submitted to the IEEE for possible publication. Copyright may be transferred without notice, after which this version maybe no longer be accessible.}}
	
	\markboth{}%
	{Shell \MakeLowercase{\textit{et al.}}: Bare Demo of IEEEtran.cls for IEEE Journals}
	
	\maketitle

	\begin{abstract}
Objective: Many recent studies have suggested that brain deformation resulting from a head impact is linked to the corresponding clinical outcome, such as mild traumatic brain injury (mTBI). Even though several finite element (FE) head models have been developed and validated to calculate brain deformation based on impact kinematics, the clinical application of these FE head models is limited due to the time-consuming nature of FE simulations. This work aims to accelerate the process of brain deformation calculation and thus improve the potential for clinical applications. Methods: We propose a deep learning head model with a five-layer deep neural network and feature engineering, and trained and tested the model on 1803 total head impacts from a combination of head model simulations and on-field college football and mixed martial arts impacts. Results: The proposed deep learning head model can calculate the maximum principal strain for every element in the entire brain in less than 0.001s (with an average root mean squared error of 0.025, and with a standard deviation of 0.002 over twenty repeats with random data partition and model initialization). The contributions of various features to the predictive power of the model were investigated, and it was noted that the features based on angular acceleration were found to be more predictive than the features based on angular velocity. Conclusion: Trained using the dataset of 1803 head impacts, this model can be applied to various sports in the calculation of brain strain with accuracy, and its applicability can even further be extended by incorporating data from other types of head impacts. Significance: In addition to the potential clinical application in real-time brain deformation monitoring, this model will help researchers estimate the brain strain from a large number of head impacts more efficiently than using FE models. 
	\end{abstract}
	
	
	\IEEEpeerreviewmaketitle

	\section{Introduction}
	Affecting over 40 million children and adults worldwide, mild traumatic brain injury (mTBI, Table 1 listed the acronyms/abbreviations used in this paper and their respective meanings), has become a serious global health challenge \cite{Cassidy}. Evidence has suggested that mTBI can lead to unconsciousness immediately after a head impact and can further result in post-concussive symptoms including cognitive deficits and emotional challenges \cite{Wallace, Dan} and even increase the risk of long-term neurodegenerative diseases such as Parkinson's and Alzheimer's diseases \cite{Doherty,guskiewicz2007}. mTBI is not simply limited to vehicle accidents, but also occurs frequently in sports such as football \cite{Montenigro17, Mez19} and mixed martial arts (MMA) \cite{MMA}. In sports-related mTBI, brain damage can accumulate with repetitive injuries, leading to more severe consequences \cite{Repeat,Tagge18,Saboori19,Hardy01,OKeeffe19,guskiewicz2003}.
	
	The fast diagnosis and early warning approaches of mTBI \cite{beckwith2013} are crucial to helping prevent repetitive sport-related mTBI since the in-time intervention after early detection can attenuate the injury to a significant extent \cite{ponsford2001,guiza2017}. As suggested in the previous work \cite{Wu19}, brain strain is a promising mechanical parameter to predict the risk of mTBI and can be useful information for medical professionals in the diagnosis of mTBI. Currently, several finite element (FE) models have been validated to calculate brain strain \cite{Zacharaki08,Saboori19,Gomez19,garimella2019,kang1997,zhao2019}; However, expertise of biomechanics is needed to perform the FE simulations, and the simulations usually take at least several hours or even days to run with a personal computer. Limited by the time-consuming nature of the FE simulations and the lack of access to computational resources, the brain strain is not widely referenced by medical professionals in the diagnosis of mTBI. Therefore, a fast and easy-to-use approach to calculate brain strain is needed for more practical clinical applications.

	As the input necessary to calculate the brain strain resulting from a head impact, the head kinematics of the impact are traditionally measured by wearable sensor systems like the head impact telemetry system (HITS)\cite{HITS}, Xpatch\cite{X2} and instrumented mouthguard\cite{MGValidation,camarillo2013,king2015, patton2016, birmingham2013}. Recently, the brain strain calculated based on mouthguard-measured head kinematics was found to correlate well with resulting mTBI \cite{OKeeffe19,Hernandez15,BAM}, which suggests the feasibility to using instrumented mouthguards to help diagnose mTBI with brain strain. Therefore, a promising approach to detect mTBI is to combine the wearable sensor systems with a fast method to calculate brain strain, and use the brain strain to predict mTBI. This approach will provide the calculated brain strain in real-time to medical professionals to assist in diagnosis and therapeutic options.
	
	Researchers have recently made efforts to develop fast and accurate methods to calculate brain strain \cite{Wu19b, BAM, Panzer1, Panzer2, atlas}. One solution is to simplify the mechanical characteristics of the brain as reduced-order brain models\cite{BAM, Panzer1, Panzer2}. As a trade-off, the reduced-order models are limited in accuracy and can only predict a single value for the brain deformation throughout the entire brain. Another potential solution to accelerate the calculation is to use machine learning tools \cite{Wu19b}. Over the last decade, machine learning has shown to be very effective in applications such as biomedical signal processing leading to improved decision making and physics-based modeling of biological and biomedical systems \cite{Zhan18, Zhan20, Wang20,Wu19b,alber2019}. Recently, as the first step to address the problem of mTBI prediction with machine learning, Wu et al. \cite{Wu19b} applied a convolutional neural network (CNN) to predict the $\mathrm{95}^\mathrm{th}$ percentile maximum principal strain (95\% MPS) of the entire brain and the corpus callosum, as well as the $\mathrm{95}^\mathrm{th}$ percentile fiber strain of the corpus callosum. This work presented a new idea: the head angular velocity profiles can be translated into images, which were the input to the CNN model to predict the resulting brain strain. The promising accuracy showed that the CNN model could be effectively applied to analyze on-field head impact for the prediction of 95\% MPS, 95\% MPS and 95\% fiber strain at corpus callosum. However, only three strain values were provided to indicate the severity of deformation, while the distribution of the deformation severity over the whole brain, which would be helpful for diagnosis, was not predicted.

	To address the costly computation of FE simulations and also provide the MPS of the entire brain, we developed a deep learning head model to calculate the peak MPS of every element of the brain. We used the kinematic data (angular acceleration and angular velocity) of 1422 impacts generated by simulations of FE model of hybrid III anthropomorphic test dummy (ATD) impacts and 381 impacts collected from on-field football and MMA games with a previously validated instrumented mouthguard. The KTH head model was used, which is a validated and widely used FE model\cite{OKeeffe19,Hernandez15,Kleiven07, Kleiven14, Kleiven13}, to calculate the true brain MPS. Engineered features representing the time-domain information of the kinematic data were extracted, and a deep neural network (DNN) was developed to predict the MPS at each brain element. To obtain datasets that were large enough to train the model, the 1422 impacts from simulations were used as the basis dataset. Then, the basis dataset was fused with the on-field impacts data to apply our model to different sports. We trained and tested our models on each of the datasets and finalized a model trained on the mixture of all datasets. Examining several metrics, the prediction of entire brain MPS was shown to be accurate. Furthermore, we found that the 95\% MPS calculated based on the entire brain MPS was comparable to the previously reported models \cite{Panzer2,Wu19b}, which showed the applicability of machine learning as a novel approach in the field of fast brain deformation calculation.

	\section{Methods}\label{sec:methods}

	\begin{table}
		\centering
		\caption{The acronyms and abbreviations used in this paper and their respective meanings.}
		\label{tab:1}       
		\begin{tabular}{cc}
			\hline\noalign{\smallskip}
			Acronym/Abbreviation & Meaning  \\
			\noalign{\smallskip}\hline\noalign{\smallskip}
	        mTBI & mild traumatic brain injury \\
	        MMA & mixed martial arts \\
	        FE & finite element \\
	        HITS & head impact telemetry system \\
	        CNN & convolutional neural network \\
	        MPS & maximum principal strain \\ 
	        ATD & anthropomorphic test dummy \\
	        CF1 & college football dataset 1 \\
	        CF2 & college football dataset 2 \\
	        HM & ATD head model simulated dataset \\
	        DNN & deep neural network \\
	        ReLu & rectified linear unit \\
	        Adam & adaptive moment estimation \\
	        MAE &  mean absolute error \\
	        RMSE & root mean squared error \\
	        Spearman Corr. & Spearman coefficient of correlation \\
	        STD & standard deviation \\
	        CI & confidence interval \\
	        RNN & recurrent neural network \\

			\hline\noalign{\smallskip}
		\end{tabular}
	\end{table}
	
	\subsection{Data description}

    To broaden the applicability of the model, we used the data of the head impacts collected on-field in college football and MMA. Instrumented mouthguards, which have been validated to measure the rigid-body movement of the head accurately, were used to collect the kinematics data \cite{camarillo2013,MGValidation}. All the impacts were video confirmed. Specifically, we included 184 head impacts in college football games \cite{Hernandez15} collected by the original version of Stanford instrumented mouthguard \cite{camarillo2013} (dataset college football 1, CF1), and 118 head impacts collected in college football games by the updated Stanford instrumented mouthguard \cite{MGValidation} (dataset college football 2, CF2). We also included 79 head impacts in MMA collected by the updated Stanford instrumented mouthguard \cite{OKeeffe19} (dataset MMA). To enlarge the training dataset, we performed simulations of head impacts using a validated FE head model of the hybrid III ATD \cite{DummyFEA} (dataset head model, HM). The bare dummy head was impacted at different locations with the velocities ranging from 2 m/s to 8 m/s, and in total 1065 head impact kinematics were obtained. To further enlarge the training dataset, the kinematics in the Y and Z axes of the HM head kinematics were switched.
	
	The KTH head model \cite{Kleiven07}, a validated FE model for mTBI-level head impacts, was used to calculate the brain peak MPS based on the input head kinematics. In the simulations, the skull of the model was rigid and moved according to the input kinematics. The peaks of angular velocity magnitude (Fig. 1A) and angular acceleration magnitude (Fig. 1B) were plotted against the resulting 95\% MPS for each dataset. All the simulations using the on-field data succeeded. However, only 1422 head impacts in HM dataset succeeded because some of the simulations failed due to the high-frequency components in the simulated kinematics. Therefore, 1803 head impacts in total were used in this study. For reference, among the four datasets, the highest 95\% MPS were: HM: 0.4423, CF1: 0.5093, CF2: 0.4184, MMA: 0.7051.

    \begin{figure}
		\includegraphics[width=\linewidth]{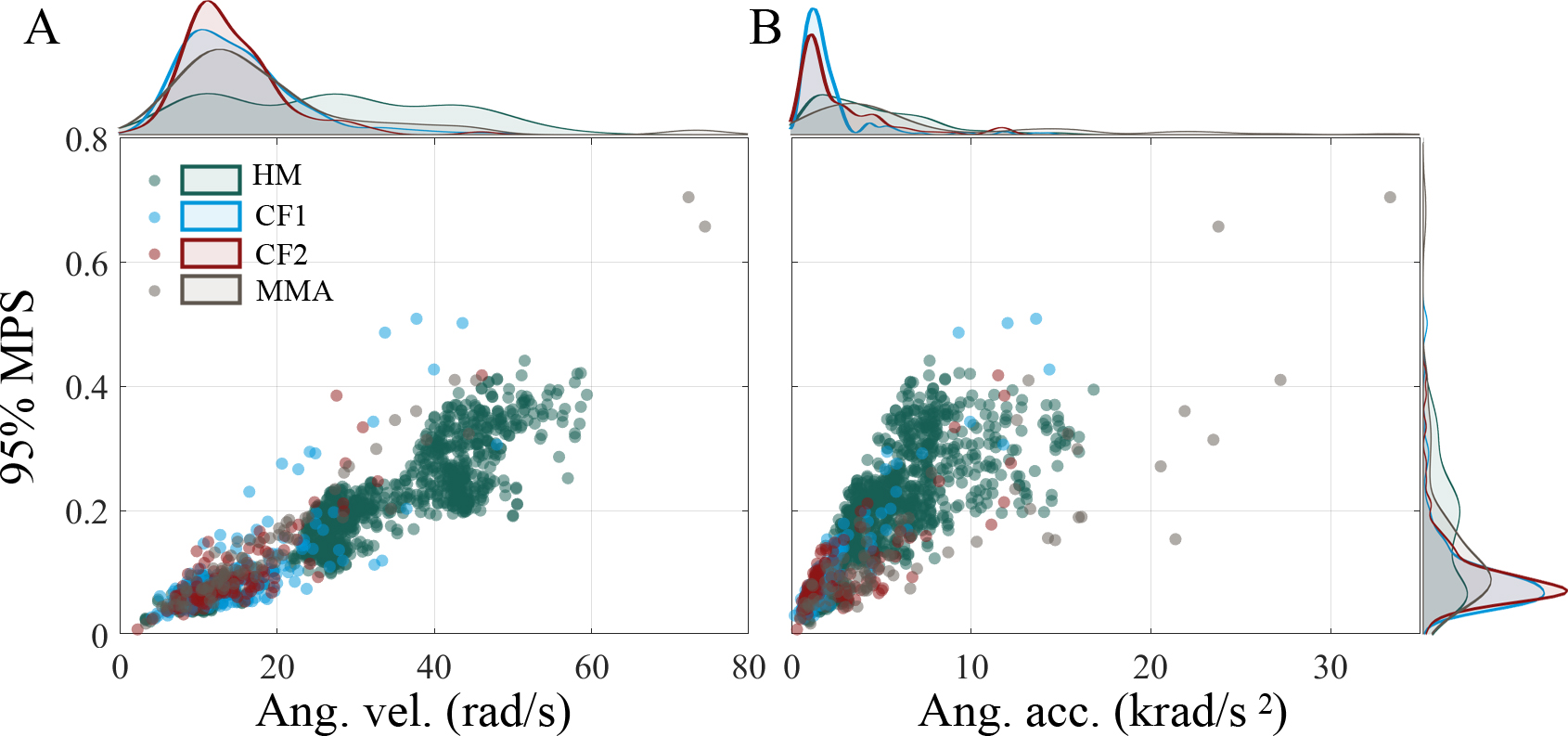}
		\caption{\textbf{Head kinematics and 95\% MPS in the datasets.} (A) 95\% MPS and peak of resultant angular (Ang.) velocity (vel.). (B) 95\% MPS and peak of resultant angular (Ang.) Acceleration (acc.). Different colors within the scatter plots indicate the different datasets (HM: hybrid III ATD impact simulations; CF1: college football impacts from the original version of instrumented mouthguard; CF2: college football impacts from the updated version of instrumented mouthguard; MMA: MMA head impacts from the updated version of instrumtented mouthguard). The curves at the top and the right are kernel density plots (Top of (A) of angular velocity. The top of panel (B) is for angular acceleration, and right of panel (B) is for 95\% MPS).}
	\end{figure}
	
	\subsection{Feature engineering and data preprocessing}
	
    For each impact, features describing the characteristics of rotational kinematics were extracted as the input to the model. Angular accelerations ($\mathbf{\alpha_{x}},\mathbf{\alpha_{y}},\mathbf{\alpha_{z}}$) were calculated by 5-points stencil derivative on angular velocities ($\mathbf{\omega_{x}},\mathbf{\omega_{y}},\mathbf{\omega_{z}}$). Then the magnitudes ($|\mathbf{\alpha}|, |\mathbf{\omega}|$) and components in each direction ($x, y, z$) of angular accelerations and angular velocities were defined as 8 channels (denoted as $c_{i}(t), i=1,2,...,8$). For each channel, six types of time-domain features were extracted based on the rationale of including the information of the signal intensity and time history:
    
    \begin{enumerate}
    	\item Maximum value (1 feature/channel): ${\mathrm{max}(c_{i}(t))}, i=1,2,...,8$
    	\item Minimum value (1 feature/channel): ${\mathrm{min}(c_{i}(t))}, i=1,2,...,8$
    	\item Integral of the time-signal (1 feature/channel): ${\mathrm{\int} c_{i}(t)dt}, i=1,2,...,8$
    	\item Integral of the absolute values of time-signal (1 feature/channel): ${\mathrm{\int} |c_{i}(t)|dt}, i=1,2,...,8$
    	\item Maximum and minimum of the exponential moving average of the signal derivative (6 features/channel): 
    	\begin{equation}
    	E_{a,i}=\{\mathrm{min}(y_{i}(n));\mathrm{max}(y_{i}(n))\}, n = 0, 1, 2, ...
    	\end{equation}
    	$E_{a,i}$ was extracted to take into account the variation of kinematic signals because it describes the general trend of a curve without being biased by extreme values caused by noise. The discrete exponential moving average of derivative $y_{i}(n)$ was defined as:
    \begin{equation}
    y_{i}(0)=ac_{i}(0)
    \end{equation}
    \begin{equation}
    y_{i}(n)=(1-a)y_{i}(n-1)+a(c_{i}(n)-c_{i}(n-1)), n \ge 1
    \end{equation} 

    Where $a$ is the smoothing coefficient chosen to be $\frac{1}{SR}$, $\frac{1}{10*SR}$, $\frac{1}{100*SR}$ to account for different smoothing effects \cite{Zhan20} and $SR$ is the sampling frequency (1kHz). $E_{a,i}$ contains the minimum and maximum values of the exponential averages of each of the derivatives of the 8 channels, and reflects the largest positive slope and the largest negative slope in the signal.
    	
    	\item Information of extrema except max and min (10 features/channel): the total number of positive extrema, the total number of negative extrema, the values of the second-largest to the fifth-largest positive extrema, the values of the second-smallest to the fifth-smallest negative extrema. This type of feature was extracted because oscillatory impacts might have an accumulating effect on the brain deformation.
    \end{enumerate}
    
    After feature engineering, the kinematics of each impact was transformed into a numerical matrix with 160 columns (8 channels $\times$ 20 features/channel), while each column denoted one feature and each row denoted one impact. The feature matrix was then used to train the deep neural network after standardization and data augmentation. Data standardization was performed based on the following formula to eliminate the bias caused by different units and imbalanced weights of feature values:

	\begin{equation}
	\bar{f}(x,i) = \frac{f(x,i)-\mathrm{mean}_x(feature(x,i))}{\mathrm{std}_x(feature(x,i))}
	\end{equation}
	Here, $f(x,i) $ denotes the $i^{\text{th}}$ feature of the $x^{\text{th}}$ impact. $\mathrm{mean}_x()$ and $\mathrm{std}_x()$ calculates the mean value and the standard deviation over the variable $x$, respectively. The standardization parameters were calculated based on the training set in the validation process, and on both the training and validation sets in the model evaluation process (See details in Section 2D).
	
	Furthermore, to optimize model performances, two commonly-used data transformations of the numerical values of labels (MPS) were done :
	
	1) Logarithmic transformation was firstly performed to stabilize the error variance and prevent negative predictions of MPS. In the training process, each MPS value was transformed by taking its logarithm. After prediction, the predicted values were inverse-transformed as MPS by exponentiating.
	
	2) The logarithmic-transformed labels were then whitened to address the problem of potential imbalanced weights given to different MPS values and remove the underlying correlation among features to improve accuracy. The whitening process was based on the following formula:
	
	\begin{equation}
	z_{whitened}(x,i) = \frac{z_{raw}(x,i)-\mathrm{mean}_x(z_{raw}(x,i))}{\mathrm{std}_x(z_{raw}(x,i))}
	\end{equation}
	
	Where $z_{raw}(x,i)$ denotes the MPS value of the $i^{\text{th}}$ element for the $x^{\text{th}}$ impact. The $\mathrm{mean}_x(z_{raw}(x,i))$ and $\mathrm{std}_x(z_{raw}(x,i))$ for each of the 4124 elements were calculated on the training set in the model validation process, and on both the training and validation sets in the model evaluation process (See details in Section 2C, 2D). $\mathrm{mean}_x(z_{raw}(x,i))$ and $\mathrm{std}_x(z_{raw}(x,i))$ were recorded to inverse-transform the prediction of MPS. The assumption was that the dataset used to calculate $\mathrm{mean}_x(z_{raw}(x,i))$ and $\mathrm{std}_x(z_{raw}(x,i))$ represented a general distribution. 
	
	The feature engineering was done on MATLAB R2020a (Austin, TX, USA). The data standardization and transforms were done on Python 3.7 with scikit-learn packages \cite{sklearn}. 
	
	\subsection{Deep learning model development}
    The goal of this work was to use a deep neural network (DNN) to act as a function approximator to learn the relationship between the measured input kinematics and the calculated MPS from FE simulations. The DNN would act as a proxy for traditional FE simulations in order to speed-up workflow when calculating MPS based on input head kinematics. To do this, the dataset must be partitioned into a training set (to train the DNN-based model), validation set (to tune hyperparameters) and test set (to evaluate model performance). Once trained, this network would take features for a new impact that was not present in the training set and the validation set and predict the MPS. 
    
    Our deep neural network consisted of five layers in addition to the input layer with 160 units and the output layer of 4124 units: 1) Hidden layer 1: 300 neurons with the rectified linear unit (ReLU) as the activation; 2) Dropout layer 1 with a dropout rate of 0.5 and no activation; 3) Hidden layer 2: 100 neurons with ReLU as the activation; 4) Dropout layer 2 with a dropout rate of 0.5 and no activation; 5) Hidden layer 3: 20 neurons with ReLU as the activation. The general design rationale was to condense kinematics information into a deep, low-dimension hidden pattern and then connect to the 4124 output units. Furthermore, dropout was used as a form of regularization to boost the model robustness and generalizability. Random initialization was used to break network symmetry. L2 regularization was also added to penalize large weights to improve generalizability of the model. Mean squared error was used as the loss function, and adaptive moment estimation (Adam) optimizer was used as the optimizer \cite{adam} with a batch size of 128 to boost the training efficiency. 
    
    On the validation set, the hyperparameters tuned in the modeling included the number of training epochs, the learning rate, and the strength of the L2 regularization. The optimal hyperparameters were chosen based on the learning curves of the training loss and the validation loss on the validation set.
    
    Before training, in an attempt to further boost the generalizability of the model, the training set was augmented by adding Gaussian noise into the training samples. A normal distribution with a mean of zero and standard deviations of 0.01 and 0.02 times the standard deviation of the original data were added. Ultimately, after the completion of this data augmentation, the number of impacts in the training set was tripled. Data augmentation and deep learning were performed in Python 3.7, using the Keras package with the TensorFlow 2.0 backend \cite{keras}.

	\subsection{Model assessment}
	The model training and assessment processes are shown in Fig. 2. Three types of tasks were performed to assess the model. Firstly, the model was trained and tested solely on the HM dataset because of its large number of impacts (Task 1). Secondly, to apply the model to specific sports, the HM dataset was used as the basis of the training dataset. The on-field data was partitioned into two parts. One part was fused into the training set and the validation set, and the rest was used for the testing (Task 2). Finally, the HM and all the on-field datasets were combined into one mixed training/validation/test set to test the model's applicability to both football and MMA impacts as a whole (Task 3). The partition of the datasets in these three tasks was the following:
	
	1)Basis: we firstly developed a model solely based on the HM dataset, which was partitioned into a training, validation, and test set (70\%, 15\%, 15\%, respectively). The entire process was repeated in 20 times with random dataset partitions and random model initialization to ensure the reproducibility of the results. 
	
	2)On-field: to balance the ratio of the basis dataset and on-field datasets in training, we fused the HM data with 80\% of the CF1/CF2/MMA data to predict the remaining 20\% of the CF1/CF2/MMA data in a 5-fold cross-validation model. In each fold, 20\% of the CF1/CF2/MMA data were used as the test set, 20\% of the remaining 80\% of the CF1/CF2/MMA data were used as the validation set. After performing 5-fold tests, all of the impacts in the on-field datasets were predicted in the test set once. 
	
	3)Mixture: we directly concatenated all the impact feature vectors of all the datasets and performed a dataset partition of training, validation and test set (70\%, 15\%, 15\%) and followed the same pattern as in the basis task with 20 repeats with randomness. 
	
	\begin{figure}
		\includegraphics[width=\linewidth]{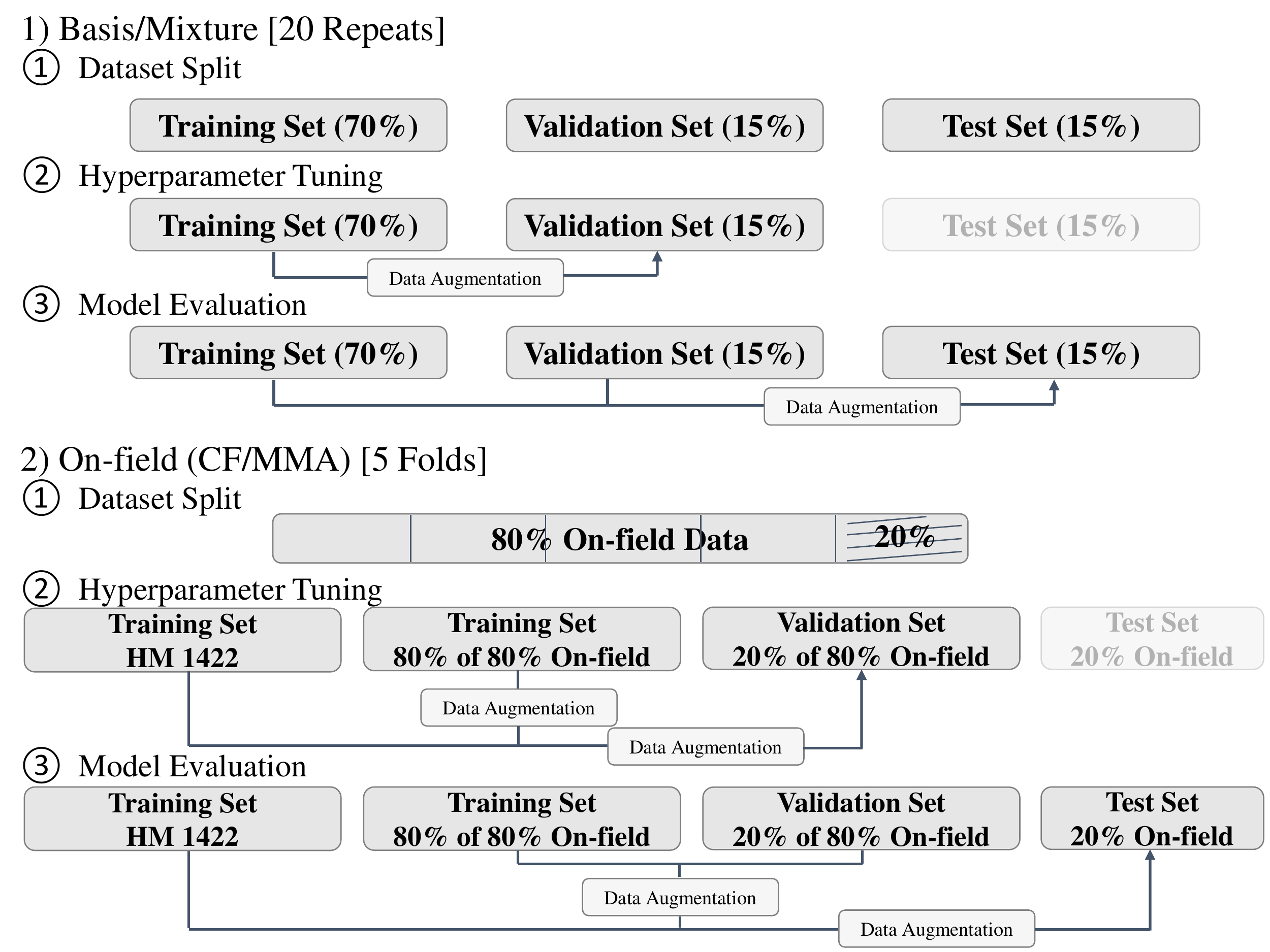}
		\caption{\textbf{The description of the dataset partition process and the function of each dataset.} In each of the three tasks (basis, on-field, and mixture), different training sets, validation sets and test sets were built. The validation sets were used to tune the hyperparameters while the test sets were used to evaluate model performances in unseen data. Data augmentation with Gaussian noise was used to improve the model generalizability. To test reproducible performances and model robustness, in the basis and mixture tasks, the experiment was done within 20 repeats with random data partitions and random model initialization, and, in on-field tasks, five-fold tests were done with randomly shuffled on-field data and random model initialization.}
	\end{figure}
	
	In each of the datasets, we trained the model on the training set and tuned the hyperparameters based on the performances on the validation set. To evaluate the model performance, we trained the model on the combination of the training set and validation set and tested it on the test set. To deal with the imbalanced dataset caused by the majority of HM data, in the on-field tasks, the same data augmentation mentioned in Section 2C was firstly used on the on-field training data when training. 
	
    For each impact, to describe the accuracy of the prediction, mean absolute error (MAE), root mean squared error (RMSE), coefficient of determination ($R^{2}$), and Spearman coefficient of correlation were calculated by comparing the predicted MPS and the true MPS at every element. MAE and RMSE show the average error over all elements and they are in the same unit as strain. RMSE was also calculated because it is more sensitive to large errors and was thus used for hyperparameter tuning. $R^{2}$ shows the linear correlation between the predicted and the true MPS, and the Spearman coefficient of correlation represents the consistency of the rank and the non-linear correlation. 
    
    Furthermore, since 95\% MPS over all elements were used to indicate the most severe brain deformation during the head impact \cite{BAM,Panzer2,Panzer1,Wu19b}, $R^{2}$ and RMSE of 95\% MPS were calculated over the impacts in the test set. The RMSE of 95\% MPS was in the same unit as strain.
	
	The accuracy metrics mentioned above (mean values of MAE, RMSE, $R^{2}$, Spearman coefficient of correlation, 95\% MPS $R^{2}$ and 95\% MPS RMSE) were calculated by comparing the predicted and the true (calculated via the KTH model) MPS at every element for every impact in the test set. As a result, we obtained a group of metrics describing the performance of the model in each repeat or fold. The summary statistics (mean, median, STD, 95\% CI) over repeats and folds were calculated in the following manner:
	
	\begin{itemize}
	\item \textbf{MAE Mean} = mean$_r$(mean$_i$(MAE$_e$(MPS)))
	
	\item \textbf{MAE Median} = median$_{r}$(mean$_{i}$(MAE$_{e}$(MPS)))
	
	\item \textbf{MAE STD} = std$_{r}$(mean$_{i}$(MAE$_{e}$(MPS)))
	
	\item \textbf{MAE 95\%CI} = 95\%CI$_{r}$(mean$_{i}$(MAE$_{e}$(MPS)))
	
	\item \textbf{95\%MPS R2 Mean} = mean$_{r}$(R$^{2}_{i}$(95\% (MPS))
	
	\item \textbf{95\%MPS RMSE Mean} = mean$_{r}$(RMSE$_{i}$(95\% (MPS))
	
	\end{itemize}
	
	For these equations, the subscripts represent that the function is calculated over that variable; for example, the subscript $r$ denotes repeats or folds, the subscript $i$ denotes impacts in the test set in one repeat or fold, and the subscript $e$ denotes the 4124 brain elements in one impact.	
	
	\subsection{Feature analysis}
	To interpret the relative importance of the various engineered features, we input each type of feature individually and classify the 160 features according to their physical meanings (and regard them as subsets of features based on their classification). The predictive power of each subset was assessed by RMSE of the predictions when only that specific subset was fed to the deep learning model. Firstly, to investigate the predictive power of various methods to engineer features, we compared the RMSE of the predictions based on maximum values, minimum values, integral values, integral of absolute signal values, maximum and minimum of the exponential average of derivative, and information of extrema except max and min. Secondly, the predictive power contributed by the angular acceleration features and the angular velocity features were compared. Thirdly, we investigated the predictive power contributed by the features related to signal values (maximum, minimum, extrema) versus features related to the time history (integral, integral of absolute signal values, and maximum and minimum of the exponential moving average of derivative). Finally, we studied the contributions of the components of the kinematics at different directions and the magnitude of the kinematics to the predictive power. The analysis was performed on the mixture datasets with the same partition described in Section 2D. Wilcoxon signed-rank test was done to determine the statistical significance of the RMSE difference between every pair of the feature subsets because the Shapiro-Wilk test rejected the assumption of data normality at a significance level of 0.05 on the RMSE of some feature subsets.

	\section{Results}\label{sec:results}
	\subsection{Model performances on different datasets}
	 The model performances based on the accuracy metrics mentioned in Section 2D are listed in Fig. 3. Meanwhile, the number of impacts used for training, validation and testing, as well as the optimal hyperparameters used in the deep neural network, are also shown in Fig. 3. Furthermore, the absolute error of MPS at every element were averaged over impacts and repeats (folds for Task 2). The kernel density distribution is plotted in Fig. 3. 
	
    It should be noted that \textbf{MAE Mean} and \textbf{MAE Median} in all tasks were smaller than 0.03, which were smaller than the differences of brain strain seen in injury and non-injury cases\cite{Kleiven07}. Furthermore, the \textbf{$R^{2}$ of 95\% MPS} were comparable to a recently reported kinematics-based injury criteria \cite{BAM, Wu19b}, suggesting that this DNN-based model is potentially accurate enough for clinical application. 
    
	\begin{figure*}
		\includegraphics[width=\linewidth]{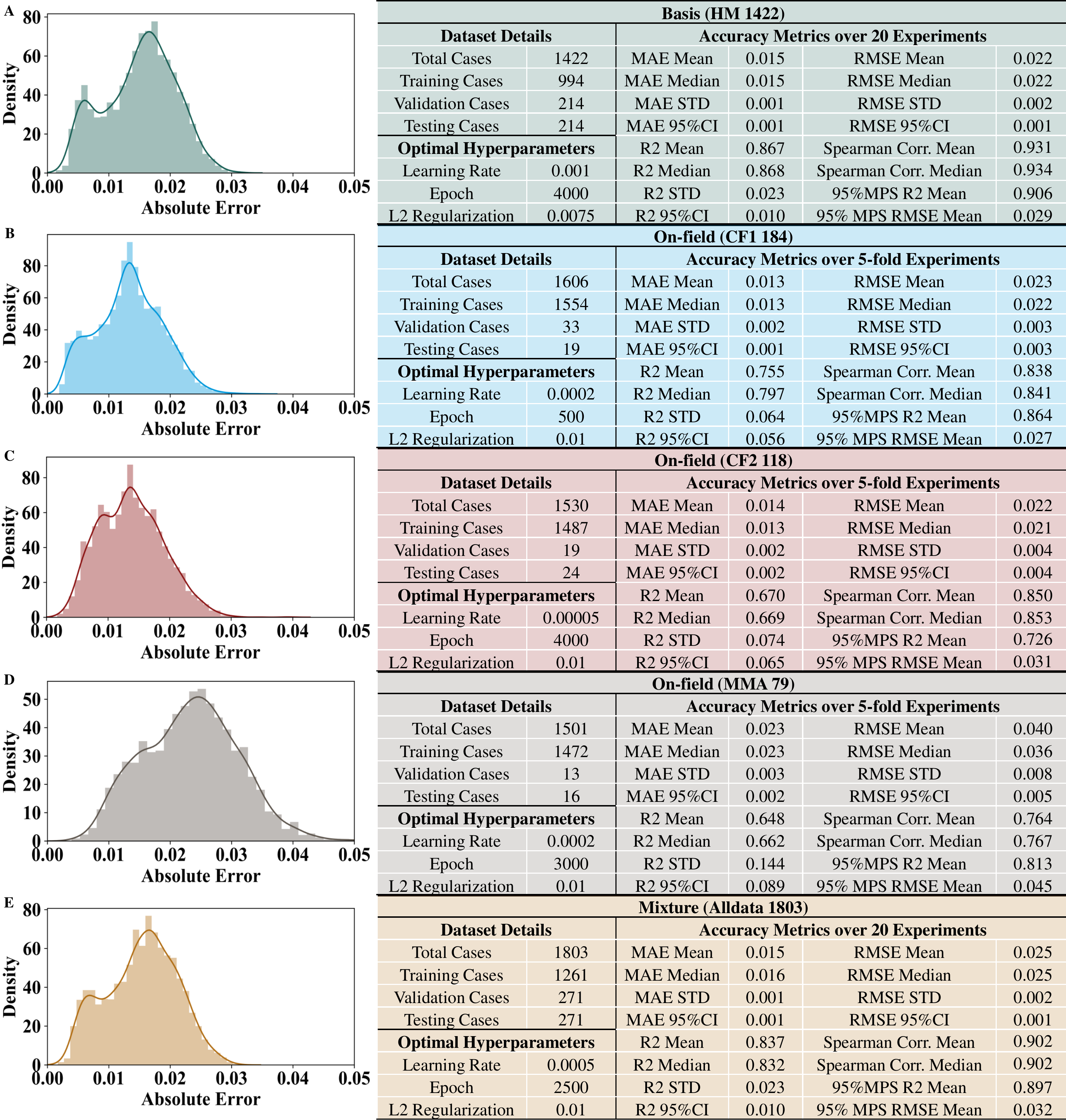}
		\caption{\textbf{The distribution of absolute error between predicted MPS and true MPS, and the model information and the accuracy metrics on different prediction tasks.} Each of the subplots shows the distribution of 4124 absolute error values calculated by averaging absolute error over test impacts and over repeats/folds for all brain elements. Each of the tables shows the data, the optimal hyperparameters used in modeling, and the model performances based on the accuracy metrics. In these tables, summary statistics over 20 repeats/5 folds are given.}
	\end{figure*}
	
	\begin{figure*}
		\includegraphics[width=\linewidth]{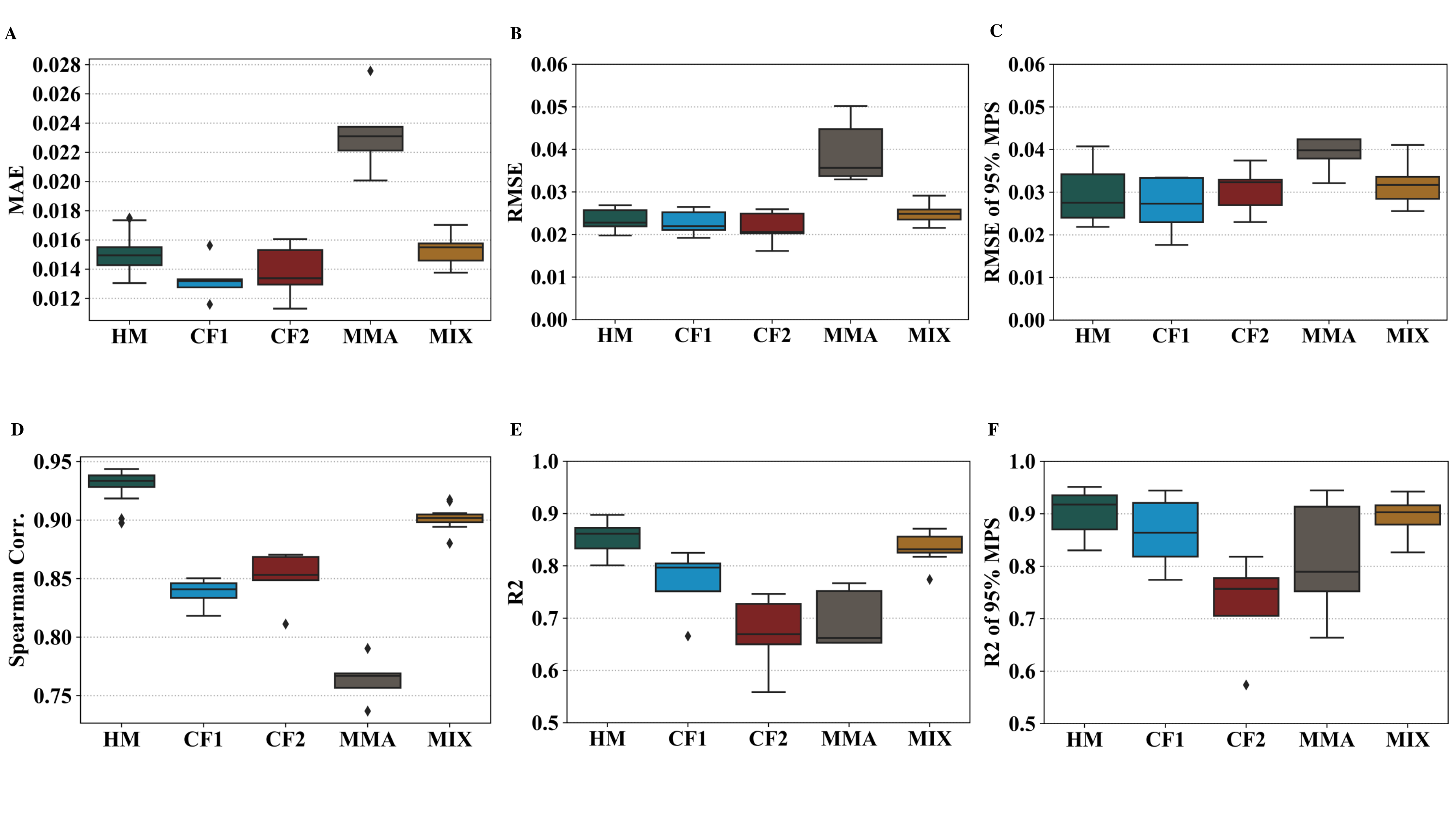}
		\caption{\textbf{Box plots of multiple accuracy metrics of models on different tasks.} Each of the subplots shows one of the accuracy metrics of models on each of the different tasks, comparing DNN predicted MPS vs. FE model calculated MPS.}
	\end{figure*}
	
	The performance of the model in different tasks is compared in Fig. 4. The predictions were more accurate on the HM and the mixture tasks than the on-field task, particularly on the MMA dataset. However, the values of mean absolute error were smaller than 0.03 for all tasks. The values of $R^{2}$ of 95\% MPS in all tasks were around or higher than 0.70 and the values of the Spearman coefficient of correlation in all tasks were around or higher than 0.75, which indicated good model accuracy in the prediction of MPS.
	
    In Fig. 5A-F, to visualize the prediction accuracy, 3 cases were selected as examples. The 3 cases were selected as the impacts with the 30th, 60th, 90th percentile highest true 95\% MPS over the test impacts of all repeats. It should be noted that the brain regions with large deformation were similar in the predicted and true results. This indicated that the current model could be used to diagnose the region of the injury, instead of only giving a scalar metric predicting the risk of concussion. The MPS of each element plotted in Fig. 5A-F were compared in Fig. 5G-I. All scatters indicating MPS lied closely to the reference line $y=x$, which showed the predictions of MPS were accurate at every element of the brain.
	
	\begin{figure*}
		\includegraphics[height=\linewidth,angle=0]{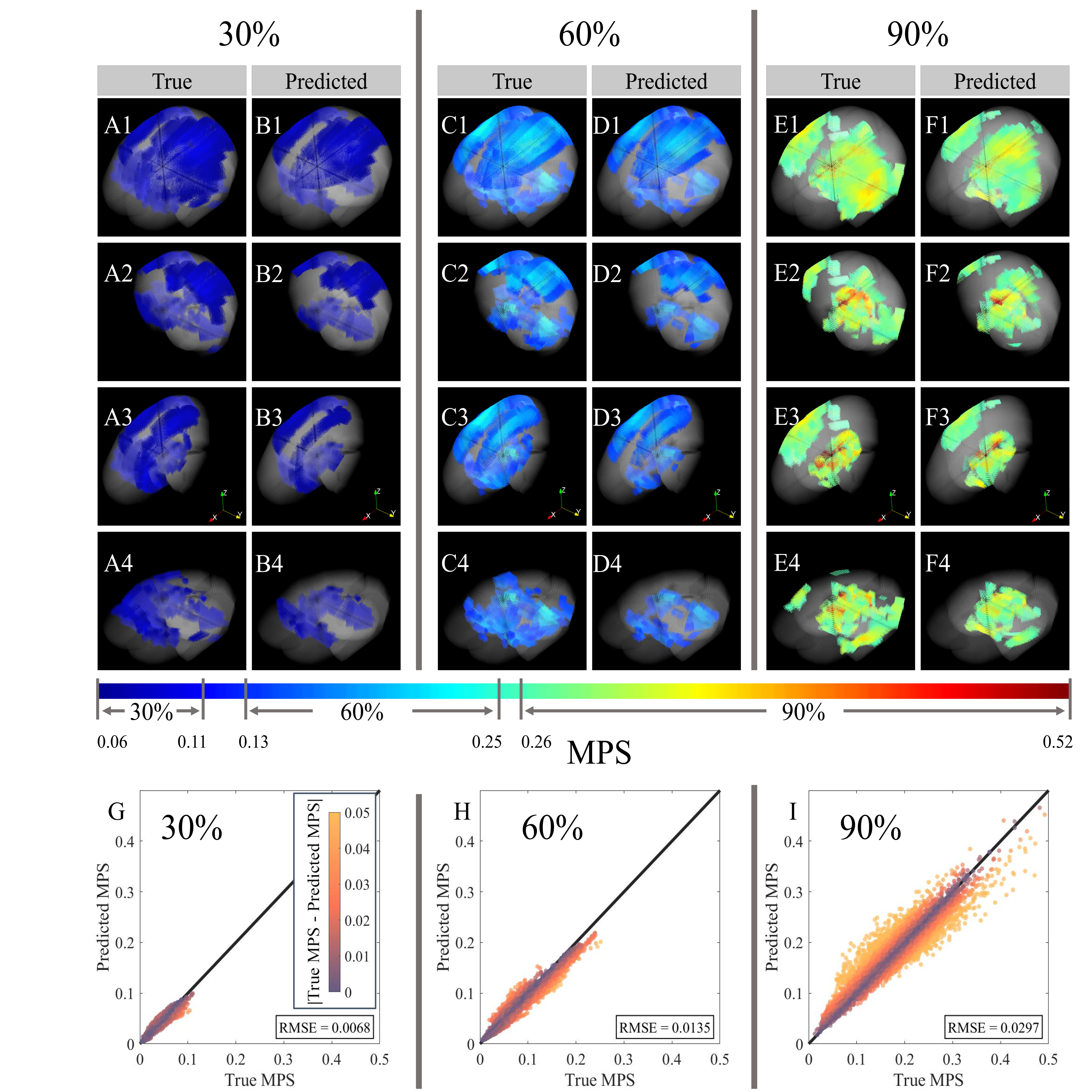}
		\caption{\textbf{The visualization of three typical predicted and true MPS in the mixture task based on 95\% MPS.} (A-F): The cloud plots of true MPS (A, C, E) and the predicted MPS (B, D, F) in the brain. From left to right, (A, B) show the 30th percentile (true 95\% MPS) case, (C, D) show the 60th percentile (true 95\% MPS) case, (E, F) show the 90th percentile (true 95\% MPS) case. From top to bottom, (A1) is the whole brain, (A2) is the posterior half of the brain, (A3) is the right half of the brain, and (A4) is the bottom half of the brain. The orientation of the brain is given in (A3). The color bar of MPS for (A-F) is given at the bottom of cloud plot, and the ranges of MPS for each case are given. The maximum and the minimum of the range are the highest MPS and half of the highest MPS, respectively. For clarity in showing the regional correlation between the true and predicted values, the elements where MPS is below the color bar range are plotted as transparent.(G-J): The comparisons between the true (KTH) MPS and predicted MPS. Each data point denotes one element in the brain. (G) is the 30th percentile case in (A, B), (H) is the 60th percentile case in (C, D), and (I) is the 90th percentile case in (E, F). The black solid line is \textit{y=x}. The colors of the scatter points represent the absolute value of the difference between true (KTH) MPS and predicted MPS. The RMSE is given at the right corner of each plot.}
	\end{figure*}
	
	\subsection{Feature analysis}
	Since, by nature, deep learning approaches tend to be a black box, we wanted to better understand what engineered features played a role in the prediction. To do so, the features were classified into several subsets according to how they were extracted from the signals (Max: maximum values, Min: minimum values, Int.: integral values, Abs. int.: integral of absolute values, EMA der.: maximum and minimum of the exponential moving average of derivative, Extr.: information of extrema except the max and min). Then, the features were also classified into subsets by the types of the kinematics that they were extracted from (Ang. acc.: angular acceleration, Ang. vel.: angular velocity); if they were extracted directly from the signal values (Value) or from the time history of the signals (His.); and if the features were calculated by the kinematics components at each direction (Comp.) or the magnitude (Mag.). The comparisons among subsets were made on the mixture dataset, and the results are shown as box plots in Fig. 6. Smaller RMSE indicates higher predictive power. The statistical significance in the RMSE difference of each pair of feature sets is shown in the four tables of Fig. 6 based on the Wilcoxon signed-rank test.

	As seen in the results in Fig. 6, in the analysis of how the features were extracted from the signals, as is expected, the model with all 160 features was the most accurate. The features from the exponential moving average of derivative exhibited significantly higher predictive power compared with other types of features($p \leq 0.001$). The maximum values, the integral values, and the integral of absolute values were found with similar predictive power ($p \geq 0.5$) and were found to be inferior to that of the EMA derivative ($p \leq 0.001$). The lowest predictive power was found in the feature subsets of minimum values and the information of extrema except max and min. In terms of the features extracted from the different kinematics, angular acceleration features showed slightly higher predictive power than the angular velocity feature ($p \leq 0.1$). This fact indicated that angular acceleration could be more predictive for MPS calculation than angular velocity based on the current feature engineering. In the comparison of features based on the signal values and the time history, signal-value related features (maximum, minimum, and extrema) were significantly less predictive when compared with the time-history related features (integral, EMA derivative, etc.) ($p \leq 0.01 $), which indicated the importance of the time history in predicting the MPS. Finally, the predictive power in the features on components of kinematics is higher than the magnitude of kinematics ($p \leq  0.01$), which agreed with the knowledge that the directional information of kinematics should be included in predicting brain deformation \cite{BAM}.
	
	\begin{figure*}
		\includegraphics[width=\linewidth]{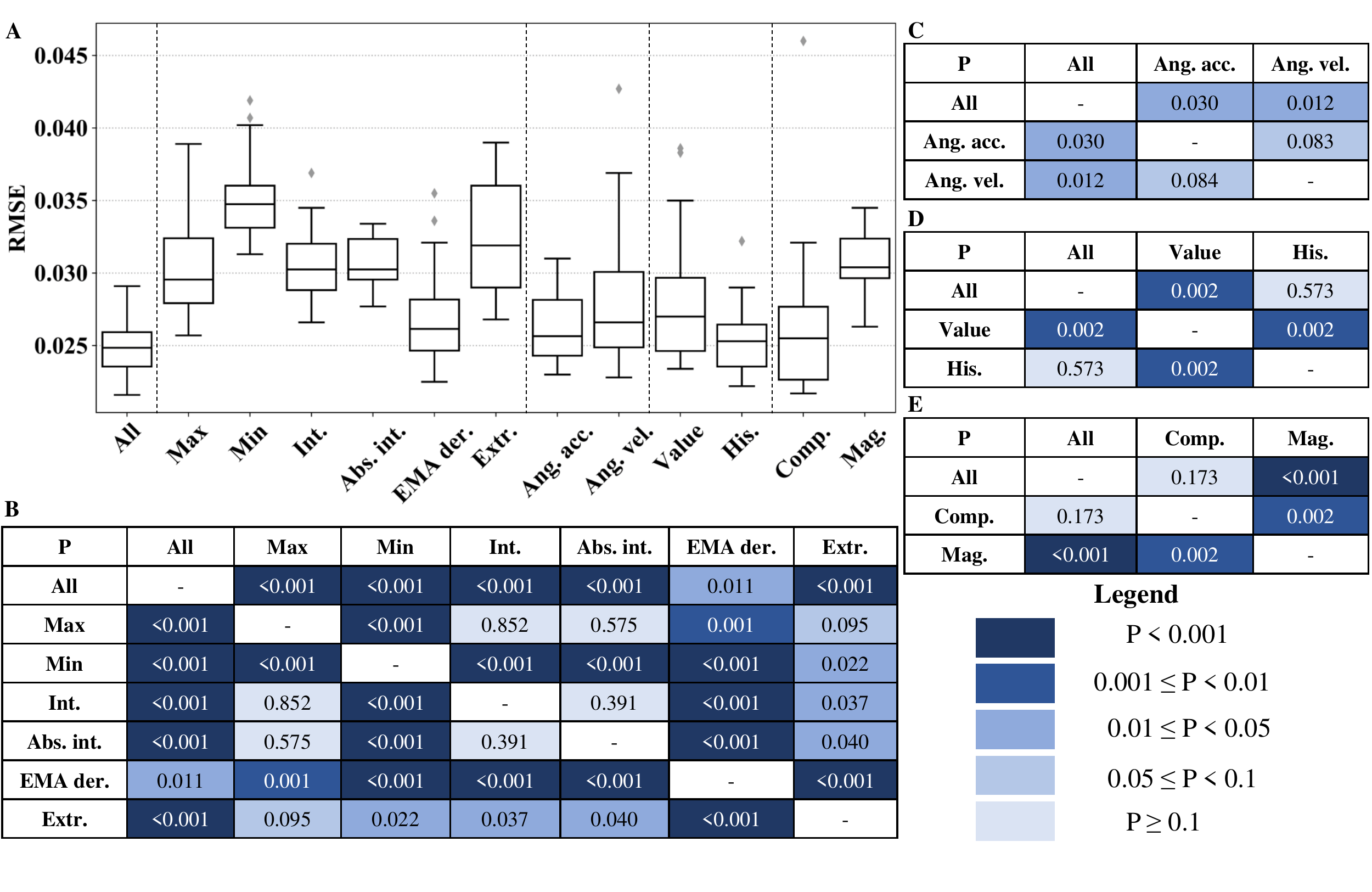}
		\caption{\textbf{The RMSE box plots of models with subsets of features and the statistical significance in the comparison between the various features based on Wilcoxon signed rank tests.} (A) The RMSE of models with different features. The dashed vertical lines separating the various box plots denote: 1) the all-feature model performance; 2) comparison among different types of engineered features; 3) comparison between angular acceleration features and angular velocity features; 4) comparison between features based on signal values and features based on signal time history; 5) comparison between kinematic components and the kinematic magnitudes.  (B, C, D) The statistical significance in comparison between pairs of features. }
	\end{figure*}	
	
	\section{Discussion}\label{sec:discussion}
	The main goal of this work is to significantly reduce the computational cost of brain strain calculation compared with conventional FE modeling. The proposed deep learning model provides users with a fast and accurate tool to calculate brain MPS in head impacts. With accuracy in MPS calculation, the major advantage of the deep learning model is the substantially smaller time consumption in the calculation: it takes $<0.001\mathrm{s}$ to calculate brain MPS using a PC (Intel Core i5-6300U). This computational time enables the real-time calculation of the brain MPS, which has promising potential to monitor the strain on the brain in real-time. As the model was trained and tested on a variety of datasets, the model can be applied to different on-field sports such as football and MMA and can be further extended to other sports with new datasets. 
	
	Previously, researchers have made great efforts to develop models to quickly estimate the brain deformation \cite{Panzer1, Panzer2, Wu19b, atlas}, and these works can be classified into two categories: reduced-order models and pre-computed models. In the reduced-order brain models \cite{Panzer1, Panzer2}, researchers modeled brain deformation using a system with mass, spring and dashpot and determined the parameters for models based on analyses of the brain response to idealized impulses. In the pre-computed models, large datasets of head impact kinematics were used to develop the atlas \cite{atlas} and train a CNN model \cite{Wu19, Wu19b}. The proposed deep learning model in our study is in the category of the pre-computed models, and the major novel development we present is that this model can predict the strain at every element of the entire brain instead of several summary values, as previously done in Wu et al., to indicate the severity of the brain deformation. The $R^{2}$ value between the predicted and the true 95\% MPS of the entire brain was 0.8970 (Mixture task, Fig. 3E), which is close to but lower than the state-of-the-art models predicting the single 95\% MPS (0.966 for the CNN model \cite{Wu19b} and 0.968 for the DAMAGE model \cite{Panzer2}). However, because the prediction types are different and the FE models used to generate the training dataset are different, the direct benchmarking against the previous models is not feasible. The performances of the models vary with datasets, as shown in Fig. 3 and also in \cite{Wu19b}. Therefore, the type of datasets (sports, measuring instrument, etc.) should be considered when interpreting the accuracy of these models.
	
	Although the proposed deep learning head model predicted the 95\% MPS with close but slightly lower $R^2$ values, this deep learning model predicted the brain strain of the entire brain with high accuracy. Firstly, the deep learning head model could predict the high-strain regions accurately based on both the visualized results and the large values of the Spearman coefficient of correlation. Because the Spearman coefficient of correlation measures the difference between the ranking of the predicted MPS and the ranking of the true MPS, the large values of the Spearman coefficient of correlation indicated that generally, the true high MPS brain elements were accurately predicted with high MPS values by the deep learning model. Secondly, the predicted MPS value on each brain element was accurate. The MAE was generally smaller than 0.03, which was smaller than the MPS difference between injury cases and non-injury cases when using the same KTH model to calculate brain strain\cite{Kleiven07}.
	
	One limitation of the pre-computed models is the dependence on the size of the training dataset. To overcome this, in a previous study, the impact kinematics data is augmented by switching the kinematics in different directions for the previously published CNN-based brain model \cite{Wu19b}. However, the rotational responses of the neck to head impacts are different in sagittal and coronal planes \cite{NeckRot}. So the augmented dataset may not accurately represent the characteristics of the head impact kinematics in football. In this study, another approach was proposed to address the limitation of the dataset size. A large dataset was generated by FE simulations of ATD head impacts (HM: 1422 impacts), and it was combined with on-field datasets (CF1: 184 impacts, CF2: 118 impacts; MMA: 79 impacts). In this method of data fusion, the deep learning model benefitted from the large number of impacts in the HM dataset. This was done by learning the mechanical characteristics from the brain model and also capturing the characteristics of the on-field kinematics in specific sports such as football and MMA. It should be noted that the FE model used in the ATD simulated impacts was originally developed for football. Since the ATD was not helmeted, the head kinematics were higher but close to helmeted head impacts because the impact impulse and neck rotation determine the head rotation. Therefore, the HM dataset might represent football style impacts better. The relatively inferior performance of the model on the MMA dataset (Fig. 3D) can be explained by this. Additionally, according to the results, it was shown that the model performed better on the HM dataset than on the on-field data. This can be explained by the lower signal-to-noise-ratio in the on-field data.
	
	Besides the new method to fuse the dataset, we also design and optimize the model structure for predicting the brain strain. The structure of the deep learning neural network applies the fully connected layers and dropout layers to extract meaningful information for the MPS prediction. The dropout layers and L2 regularization were used to avoid overfitting. As an important factor in impacting accuracy, the hyperparameters were carefully tuned based on a validation set to find the optimal hyperparameters for an accurate deep learning model with generalizability. The model performance was then tested on a separate test set with completely unseen impacts, which showed the model accuracy in predicting unseen data without overfitting to the training and validation set. To stabilize the variance and prevent large errors with large prediction values, this deep learning method applied the logarithmic transformation and data whitening, which improved the accuracy and ensured all predictions were positive because only positive MPS values have physical meanings. Additionally, the deep learning model can be more scalable when new datasets can be incorporated into the current training dataset, and re-training can be done to improve the model performance. This cannot be easily done in the less scalable pre-computed brain injury models \cite{atlas}.
	
	The feature analysis shows the contribution of each type of feature in terms of the predictive power of the deep learning model. Firstly, angular acceleration features show higher predictive power than angular velocity features. This finding contradicts the results that angular velocity magnitude was better correlated to MPS than angular acceleration in a recent study \cite{BAM}. However, it should be noted that the angular acceleration is the derivative of the angular velocity, and the features included the information of derivative and integration. So angular velocity and angular acceleration subsets might have some overlapping features. The potential reason may be that our features incorporate the time history of the signals (the integral and the integral of absolute signal values, the exponential moving average of the derivative of the signal) instead of only the maximum values, and the time history leads to the higher predictive power of angular acceleration. Further study is needed to investigate the causal relationship between angular acceleration and MPS, as well as between angular velocity and MPS. Secondly, the time history features are more predictive in general than the value features. This finding indicates the necessity of further machine learning work to make use of the temporal relationship in the data mining of kinematic signals, and also suggests the head impact measurement should focus on time history instead of just the the maximum values.
	
    The deep learning model can be applied to evaluate the risk of a head impact. FE simulation, as the conventional approach to calculate the brain MPS, takes hours or even days in the calculation. The complicated FE models also require trained engineers to perform the simulation. Using the deep learning model, medical professionals can monitor the brain MPS in a real-time manner. Furthermore, the deep learning model is simple enough to be built into a small device. With a combination of sensor signal recordings and processing in wearable devices, the risk of a head impact can be evaluated by the device, and the MPS of the entire brain can be shared with medical professionals to help diagnosis in real-time. As more data become available and as more is understood regarding the correlation of MPS and brain injury, the more realistic this becomes. Evidence suggests that the pathology occurs at the regions of high MPS \cite{OKeeffe19}, so the deep learning model can allow medical inspections (such as CT or MRI scans) to be made as soon as the calculated MPS indicates serious brain injury threats in certain brain regions. In addition, better precaution and protection decisions and plans for these players could also be put in place with this information. In the future, additional metrics like maximum principal strain rate, maximum shearing stress can also be considered to diversify the output information related to brain injury for medical analysis.
	
	Furthermore, this model benefits more than just clinicians; it can also benefit researchers working in the field of TBI and mTBI. This model provides researchers with a faster method to calculate brain MPS, which is recognized as a parameter indicating the risk of mTBI. Using the deep learning model, one head impact can be modeled in less than a second and researchers can easily process a large amount of head impacts kinematics without having to spend days and weeks working through FE simulations. 

	While this work provides a novel and powerful method for mTBI monitoring, there are some limitations beyond those previously mentioned. Firstly, this model used feature engineering to extract information from the time-domain kinematics, but the overall predictive power of the features set an upper bound on model performance. As there is no widely acknowledged standardized feature sets to characterize brain deformation, the engineered features in this work may also not be the best features to drive the proposed deep learning models. It is possible that, without artificially engineering the features, other deep learning architectures such as a convolutional neural network (CNN) and recurrent neural network (RNN), may extract features from the longitudinal signals automatically on the premise of large quantities of data. However, ideally there would be much more data in order to enact these type of models. This was shown in Wu et al. \cite{Wu19b} where a CNN was applied to estimate the MPS for the whole brain and reported accurate results. More work in this area is warranted and will continue to add better fidelity and accuracy to this new approach. Another limitation is that the deep learning model does not predict the fiber strain because the version of KTH head model used in this study does not incorporate the axonal fibers.
	
	Furthermore, although in this work high accuracy was shown when the deep learning head model was applied to different datasets, the accuracy in real-world application remains to be validated considering the different characteristics of different types of head impacts. For instance, the head impacts in high school football games and those in college football games may have different characteristics, considering the age and the level of competitiveness. Therefore, the accuracy needs to be further tested on different types of head impacts and even more features such as the age of players, can be added to improve the model accuracy.
	
	Finally, although this study endeavors to interpret the models with different input feature subsets, the information extracted in the hidden states of the model still remain to be further studied. For instance, the last hidden layer with only twenty neurons might have already condensed the information of brain deformation. As is shown in the applications in DeepFace and DeepPatient \cite{deepface, deeppatient} where the information of the deep hidden-layer neurons can be extracted for downstream pattern recognition tasks such as facial recognition and clinical outcome prediction, whether the data-driven low-dimension representation could be directly used as brain injury criteria remains an open and interesting research topic. It is possible that the information of the twenty-neuron layer can be extracted and used in prediction tasks such as concussion prediction.
	
	\section{Conclusion}\label{sec:conclusion}
	This study proposes a novel deep learning head model for fast whole-brain strain calculation. Based on the simulated head impact data and the on-field football and MMA data measured via instrumented mouthguard, with data fusion, the deep learning head model showed high accuracy, which was manifested in multiple metrics in the calculation of maximum principal strain. This model remarkably accelerates the calculation process when compared with the conventional FE simulations. The results also showed that the deep learning head model was accurate in predicting the severity of brain deformation as well as the specific regions of the brain suffering from the highest strain. Feature analysis of the model showed that certain types and groups of engineered features were more predictive in MPS prediction and may be inspirational for researchers to extract effective features and understand what factors drive MPS in future studies. This model enables researchers to calculate brain strain in a much faster and even real-time manner without performing computationally and time expensive FE simulations. Additionally, the complete, fast, and accurate brain strain visualization of the proposed model provides trainers with a reliable method to monitor the forces on the brain and the resulting brain deformation from head impacts, such that medical professionals can intervene in real-time to prevent brain damage accumulation and assist in mTBI diagnosis. 
	
	\section{Acknowledgement}\label{sec:acknowledgement}
	This work was supported by the Department of Bioengineering, Stanford University and the Office of Naval Research Young Investigator Program (N00014-16-1-2949).

\end{document}